\documentclass[fleqn,usenatbib]{mnras}

\pdfoutput=1

\usepackage[T1]{fontenc}

\DeclareRobustCommand{\VAN}[3]{#2}
\let\VANthebibliography\thebibliography
\def\thebibliography{\DeclareRobustCommand{\VAN}[3]{##3}\VANthebibliography}

\usepackage{graphicx}	% Including figure files
\usepackage{amsmath}	% Advanced maths commands
\usepackage{amssymb}	% Extra maths symbols
\usepackage{bm}         % Bold symbols in maths mode
\usepackage{ulem}       % Strikethroughs & underlines

\newcommand{\amuse}{\texttt{AMUSE}}
\newcommand{\C}{\texttt{C}} 
\newcommand{\ias}{\texttt{IAS15}}
\newcommand{\jupyter}{\texttt{Jupyter}}
\newcommand{\mercurius}{\texttt{MERCURIUS}}
\newcommand{\mesa}{\texttt{MESA}}
\newcommand{\python}{\texttt{Python}}
\newcommand{\reb}{\texttt{REBOUND}}
\newcommand{\rebx}{\texttt{REBOUNDx}}
\newcommand{\sse}{\texttt{SSE}}
\newcommand{\whfast}{\texttt{WHFast}}
\newcommand{\dts}{$dt_\textrm{split}$}
\newcommand{\Nbody}{\textit{N}-body}

\graphicspath{ {./images/} }

\title[Stellar Evolution and Tides in REBOUNDx]{Stellar Evolution and Tidal Dissipation in REBOUNDx}

\author[Stanley A. Baronett et al.]{
Stanley A. Baronett,$^{1}$\thanks{E-mail: \href{mailto:barons2@unlv.nevada.edu}{barons2@unlv.nevada.edu}}
Noah Ferich,$^{2}$\thanks{E-mail: \href{mailto:noah.ferich@colorado.edu}{noah.ferich@colorado.edu}}
Daniel Tamayo,$^{3}$\thanks{NHFP Sagan Fellow: \href{mailto:dtamayo@astro.princeton.edu}{dtamayo@astro.princeton.edu}}  
Jason H. Steffen,$^{1}$\thanks{E-mail: \href{mailto:jason.steffen@unlv.edu}{jason.steffen@unlv.edu}}
\\
$^{1}$Department of Physics \& Astronomy, University of Nevada, Las Vegas, 4505 S. Maryland Pkwy, Las Vegas 89154, USA\\
$^{2}$Department of Astrophysical \& Planetary Sciences, University of Colorado Boulder, Boulder, CO 80309, USA\\
$^{3}$Department of Astrophysical Sciences, Princeton University, Princeton, NJ 08544, USA\\
}

\date{}

\pubyear{2022}

\usepackage{newtxtext,newtxmath}
\begin{document}
\label{firstpage}
\pagerange{\pageref{firstpage}--\pageref{lastpage}}
\maketitle

\begin{abstract}
To study the post-main sequence evolution of the Solar system and exoplanetary systems more accurately and efficiently, we introduce two new features to \rebx, an extended library for the \Nbody\ integrator \reb.  The first is a convenient parameter interpolator for coupling different physics and integrators using numerical splitting schemes.  The second implements a constant time lag model for tides without evolving spins.  We demonstrate various uses of these features using stellar evolution data from \mesa\ (Modules for Experiments in Stellar Astrophysics) as an example.  The results of our tests agree with several studies in the literature on post-main sequence orbital evolution, and our convergence and performance studies respectively demonstrate our implementations' accuracy and efficiency.  These additional effects are publicly available as of \rebx's latest release.
\end{abstract}


%%%%%%%%%%%%%%%%%%%%%%%%%%%%%%%%%%%%%%%%%%%%%%%%%%

%%%%%%%%%%%%%%%%% BODY OF PAPER %%%%%%%%%%%%%%%%%%

\section{Introduction}
\label{sec:intro}

\reb\ is an open-source, modular \Nbody\ integrator, which simulates the dynamical motion of particles (e.g., stars, planets, and dust) under the influence of forces such as gravity \citep{Rein2012}.  Written entirely in \C, with memory and computational efficiency in mind, the code can also be conveniently imported as a \python\ module.  \reb\ features several integrators for calculating \Nbody\ trajectories and their derivatives \citep{Rein2016}, including \ias, a general purpose, high accuracy integrator with adaptive timesteps \citep{IAS15};  \whfast, a fast, unbiased Wisdom-Holman integrator for long-term simulations \citep{WHFAST}, as well as higher order symplectic schemes \citep{Rein2019b}; and \mercurius, a hybrid integrator based on the \texttt{MERCURY} \citep{MERCURY} algorithm to allow for close encounters \citep{Rein2019}.  Meanwhile, \rebx\ (eXtras) is an extended library and flexible framework for incorporating additional physics into its integrations, e.g., post-Newtonian corrections or radiation forces \citep{Tamayo2020}.

With the development of increasingly sophisticated codes to model different physics, leveraging numerical schemes that couple distinct integrators in a modular fashion can prove useful, particularly in the rapidly growing field of post-main sequence (MS) planetary science.  The computational challenge of full-lifetime integrations over the entire -- or even much of the -- MS, giant branch (GB), and white dwarf (WD) phases often require methods to combine \Nbody\ and stellar evolution codes and to include additional, relevant forces \citep{Veras2016a}.  For example, rather than duplicating stellar evolution models in \rebx, it would be preferable to use existing state-of-the-art codes, e.g., the open-source Modules for Experiments in Stellar Astrophysics \citep[\mesa,][]{Paxton2011, Paxton2013, Paxton2015, Paxton2018, Paxton2019}.  A simple and powerful class of schemes for coupling integrators are splitting schemes, which alternate (in this case) between evolving the orbits and the star using fixed timesteps \citep{Strang1968, Hairer2006}.  Calling integrators separately in this fashion minimizes code duplication, and the numerical scheme errors can be understood in terms of the commutation relations between the operators being combined \citep[e.g.,][]{Tamayo2020}.  This strategy has been vigorously pursued in the Astronomical Multipurpose Software Environment (\amuse) package \citep{Portegies2018book, Portegies2018, Portegies2020}, which couples a wide range of codes with splitting schemes of various orders.

While there are obvious advantages to adopting a general and widely used environment like \amuse\ for coupling \Nbody\ integrations with other astrophysical codes, a library like \rebx\ that is tailored to \reb\ allows greater flexibility for tackling challenging computational problems requiring more customized approaches.  For example, while \amuse's standard Bridge scheme would work well to apply tidal forces with the \ias\ or \whfast\ integrators in \reb, it would fail during a close encounter using the current implementation of the hybrid \mercurius\ integrator in \reb, since it would apply the forces at the end of the global timestep rather than during the close encounter when the forces are greatest.\footnote{Bridge expansions for a particular subset of particles or the Nemesis integrator in \amuse, however, may offer viable alternatives.}
By contrast, our native implementation ensures the tidal forces are appropriately included in the total force applied by \ias\ during close encounters (see \S~\ref{sec:pi_impl} and \citealt{Tamayo2020} for more details).
More broadly, the astronomical community has historically benefited from the innovation and rigor spurred by the existence of several codes that can be compared and tested against one another.

Therefore, in this paper, we develop these capabilities for the \rebx\ package.\footnote{The latest version of \rebx\ is available at \url{https://github.com/dtamayo/reboundx}.  We make available \jupyter\ notebooks and sample \python\ scripts used to generate the following results and figures at \url{https://github.com/sabaronett/REBOUNDxPaper}.  Any questions or problems can be reported by opening an issue at \url{https://github.com/sabaronett/REBOUNDxPaper/issues}.}  Most stellar evolution occurs on timescales of billions of years, while the orbits of close-in Kepler systems oscillate on timescales orders of magnitude shorter.  In this adiabatic limit, where one set of variables (e.g., stellar parameters) changes much more slowly than the other (orbital parameters), instead of evaluating stellar models from scratch at every \Nbody\ timestep, one can take the much more efficient approach of running a single stellar model and interpolating its results for a large number of \Nbody\ integrations.
To this end, we present a machine-independent implementation of parameter interpolation in \S~2 and apply it to stellar evolution data from \mesa\ as an example.
To further show the modularity of such splitting schemes, we implement a constant time lag model for tides (without evolving spins) from \citet{Hut1981} in \S~3.  We compare our examples with other works throughout this article, and we show results that combine both stellar and tidal evolution, as well as convergence and performance studies, in \S~4.

\section{Splitting Schemes for Additional Effects}
\label{sec:ss}

\subsection{Parameter Interpolation (PI) \rebx\ Implementation}
\label{sec:pi_impl}

We can couple distinct integrators that model different physics using the following numerical scheme.  Formally, we have a coupled set of differential equations for the \Nbody\ evolution $\hat N\:\mathbf{z}$ and the parameters themselves $\hat P\:\mathbf{z}$, where we define differential operators $\hat N$ and $\hat P$, which act on the current state of the system $\mathbf{z}$.  If we have a solution for the parameter differential equations in isolation, we define a corresponding integration operator $\mathcal{P}(h)$ that advances the state $\mathbf{z}$ by a timestep $h$ according to $\hat P\:\mathbf{z}$.  We can also define a solution to the \Nbody\ equations through its own corresponding integration operator $\mathcal{N}(h)$ that similarly advances the state according to $\hat N\:\mathbf{z}$.

Thus, we construct a first-order splitting scheme $\mathcal{S}$ that alternates between an \Nbody\ step for a splitting time interval and a parameter-evolution step for a splitting time interval:
\begin{equation}
 \mathcal{SNP}(dt_\textrm{split})\mathbf{z}(t) \equiv \mathcal{N}(dt_\textrm{split}) \circ \mathcal{P}(dt_\textrm{split}),
 \label{eq:splitting}
\end{equation}
where $\mathcal{N}(dt_\textrm{split})$ is made up of many \Nbody\ steps of size $dt$.  For small enough timesteps, this splitting method approximates the true solution:
\begin{equation}
  (\mathcal{N}+\mathcal{P})(dt_\textrm{split}) = \mathbf{z}(t+dt_\textrm{split}) \approx \mathcal{N}(dt_\textrm{split}) \circ \mathcal{P}(dt_\textrm{split}).
 \label{eq:exact}
\end{equation}
The integration errors of such splitting schemes can be understood precisely in terms of the non-commutative properties of the two operators \citep[see][]{Tamayo2020}.  This also helps guide an appropriate choice of \dts, such that $dt_\textrm{split} \ll \tau_\textrm{PI},$ the time-scale of the parameter evolution (see \S~\ref{sec:convergence}).

Precise adherence to this splitting scheme would use parameter integration outputs that correspond to the specific and exact time intervals of \dts.  For incorporating stellar evolution as an example, this amounts to alternating timestep calls between \reb\ and \mesa.  However, repeating runs of the same stellar model for many different \Nbody\ integrations in this way can be inefficient.

Ensuring \reb's \Nbody\ steps always fall at exactly the same times as \mesa's can be impractical for a survey of planetary systems with different orbital periods (and hence different timesteps).  This is also challenging when using integrators with adaptive timesteps as we do here and as used by \mesa.  Yet many effects, such as stellar evolution, are very slow compared to orbital time-scales.  In such adiabatic cases, a simple approach is to interpolate the results of a single \mesa\ integration at arbitrary times.  The error from interpolating at \dts, instead of evaluating $\mathcal{P}($\dts) with \mesa\ explicitly, is negligible compared to the splitting scheme error.

We introduce a new feature to \rebx\ to accomplish the following: (1) load and store parametric time-series data in the simulation's allocated memory; and (2) spline the data so users can interpolate a parameter's value at any arbitrary time in the simulation.  Using a cubic spline, we reduce the potential for Runge's phenomenon around discontinuous derivatives (compared to polynomial interpolation) when interpolating non-smooth data, e.g., stellar mass and radius profiles around the tips of the red-giant branch (TRGB) and asymptotic giant branch (AGB) (see Fig.~\ref{fig:massrad}).

We developed \rebx's interpolation functions by adapting the cubic spline algorithm from \citet{Press1992}, optimised for the \C\ language.  By adding these functions directly to the core \C\ source code, we ensure machine independence and avoid requiring users to install additional libraries or dependencies.  We also incorporate a custom and optimised searching algorithm into the interpolation function.  This function allows the code to support forward and backward integrations\footnote{E.g., using \reb's \texttt{JANUS} integrator \citep{JANUS}.} and interpolations at arbitrary times.  This `Parameter Interpolation' (PI) feature, available as of version 3.1.0, therefore allows users to import data from other codes into their \reb\ simulations.\footnote{Documentation, as well as both \C\ and \python\ examples of its uses, can be found at \url{https://reboundx.readthedocs.io/en/latest/effects.html\#parameter-interpolation}.}

The interpolator object incorporates a time series by accepting two arrays: (1) a monotonically increasing time series, in one-to-one correspondence with (2) a series of values for a given parameter.  Users can populate these arrays in any desired manner, including, but not limited to, importing values from an external data file.  For example, users can generate a discrete set of parameter values (e.g., stellar mass) from their own formulas, from their own integrations, or from existing stellar evolution codes, e.g., \mesa\ or \sse.

When using \mesa, we recommend the methodology laid out in the \texttt{mesa2txt.ipynb} \jupyter\ notebook, available at the repository for this paper (see \S~\ref{sec:intro}).  The procedure isolates a parameter from standard \mesa\ output logs and generates a two-column, tab-separated text file.  This method also accounts for when a \mesa\ integration restarts from an earlier timestep\footnote{For example, \mesa\ may automatically attempt a `backup' or `retry' when convergence fails between timesteps.} and ensures the time-series part of the data imported into \rebx\ is strictly increasing.

Before starting an integration, we create a separate interpolator object for each varying parameter.  We then repeatedly call \reb's main integration function when looping over a list of times to update the parameters to their interpolated values at each iterated time of the loop.  This results in two distinct intervals: (1) the existing integration timestep $dt$, and (2) an interpolation interval \dts.

\subsection{PI Demonstration}
\label{sec:pi_demo}

Here we interpolate stellar evolution data to demonstrate the splitting scheme in action.  As a simple example, we use \mesa\ to model the Sun's evolution from pre-MS to WD\footnote{Release version 12778, and \texttt{MESA SDK} version 20.3.2 (DOI 10.5281/zenodo.3706650)}.  \mesa\ is an open-source and modular code, capable of solving 1D stellar evolution in a wide range of environments.  Its advantages include up-to-date, independently usable microphysics modules; a fully-coupled solution for composition and abundances; advanced adaptive mesh refinement techniques; and increased performance through effective parallelism on multi-core architectures.

\mesa\ also supports various preloaded and custom mass-loss rate configurations along different evolutionary stages \citep[p. 16]{Paxton2011}.  For the red-giant branch (RGB) phase, we used the default \citet{Reimers1975} formula for \mesa's `cool-wind RGB scheme':
\begin{equation}
 \dot M = 4 \times 10^{-13} \eta \dfrac{LR}{M},
 \label{eq:reimers}
\end{equation}
where $L$, $R$, and $M$ respectively are the stellar luminosity, radius, and mass (all in solar units), and $\eta$ is a dimensionless scaling factor.  From \citet{Veras2012} and {\citet{Veras2016b}}, $0.2 \leq \eta \leq 0.8$ is a realistic range for the Sun.

\begin{figure}
 \includegraphics[width=\columnwidth]{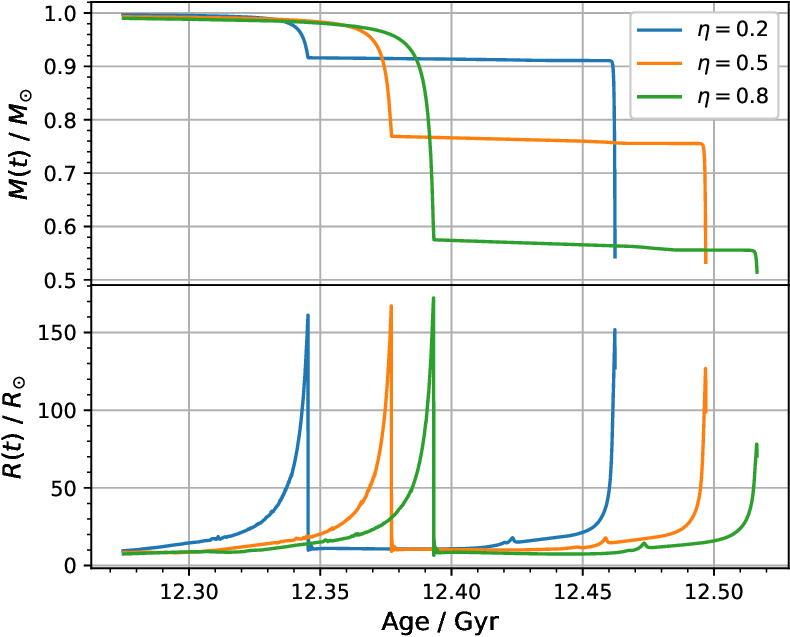}
 \caption{Post-MS mass (top) and radius (bottom) evolution of a Sun-like star, with various Reimers scaling factors (Eq.~\ref{eq:reimers}), from \mesa.  At about two-thirds in age from each TRGB to TAGB, tiny drops in mass and transient bumps in radius correspond to thermal pulses along the AGB resolved by \mesa.  For reference, 1~au $\approx 215~R_{\sun}$.}
 \label{fig:massrad}
\end{figure}

As a basic template for the following examples, we use the prepackaged `1M\_pre\_ms\_to\_wd' stellar model, part of the verification test suite included in \mesa\ \citep[][]{Paxton2011}, but a more comprehensive set of stellar evolutionary tracks are detailed in \citet[]{Choi2016}.\footnote{Our \mesa\ inlists, with modifications only to Reimers' scaling factor and which data are outputted, can be found at this paper's companion GitHub repository (URL in the ``Data Availability'' section at the end).}
Fig.~\ref{fig:massrad} shows post-MS results from \mesa\ for the mass and radius evolution of a $1~M_{\sun}$ star.
As in \citet[Fig. 1]{Veras2012} and \citet[Fig. 1]{Veras2016b}, we see a trade off between the realistic bounds of Reimers' scaling factor in the amount of mass lost by the end of the RGB and AGB phases; meanwhile, $\eta = 0.5$ finds roughly equal amounts of mass lost during each of the RGB and AGB phases.  We attribute any precise differences in RGB and AGB evolution, between our Fig.~\ref{fig:massrad} and those of \citet{Veras2012} and \citet{Veras2016b} to: (1) the different AGB mass-loss prescriptions used -- e.g., we use \citet{Blocker1995}, \mesa's default, while they use \citet{Vassiliadis1993}; and (2) the different stellar codes used -- both \citet{Veras2012} and \citet{Veras2016b} use \sse.

As seen in Fig.~\ref{fig:massrad}, and as noted in \citet[\S~2]{Schroder2008}, the precise solar mass-loss directly affects the giant Sun's radius, since reduced gravity allows for more extended, cooler giants.  In fact, their paper uses a different mass-loss prescription from \citet[Eq.~4]{Schroder2005}, which seeks to revise the original \citet{Reimers1975} law;\footnote{Despite improved agreement with observed mass-loss rates for supergiants with very low gravity, this new formula is not valid for stars like the Sun, as noted explicitly in the conclusions of \citet[\S~4]{Schroder2005}.} they also use a much older stellar evolution code, \citet{Eggleton1971, Eggleton1972, Eggleton1973}.  These differences in method account for discrepancies (e.g., TRGB age) between our \mesa\ results in Fig.~\ref{fig:massrad} and those of \citet{Schroder2008}.  Thus, long-term solar evolution results like these are sensitive to different mass-loss prescriptions and stellar evolution codes, but further analysis in this regard is beyond the scope of this article.  Furthermore, as we mention in \S~\ref{sec:pi_impl}, we do not advocate -- nor does PI restrict -- the use of any specific stellar code.

For the $\eta=0.8$ track, which shows the most RGB-mass loss, we twice simulate an idealised Sun-Earth system roughly 4 million years (Myr) before the TRGB using \whfast.\footnote{A \jupyter\ Notebook of this interpolation example can be found at \url{https://github.com/dtamayo/reboundx/blob/master/ipython_examples/ParameterInterpolation.ipynb}.}  We invoke our new PI code in \rebx\ to load in the Sun's post-MS \mesa\ data to interpolate and update its mass and radius.  We do this first with \dts$=4000$~yr and second with \dts$=400$~yr (a 10x-shorter interval) to observe the numerical splitting scheme's trajectory in terms of the quantitative results.  We initialise Earth's semi-major axis at 1~au, although in reality its orbit would have expanded somewhat from any stellar mass loss prior to the start of our simulation.

\begin{figure}
 \includegraphics[width=\columnwidth]{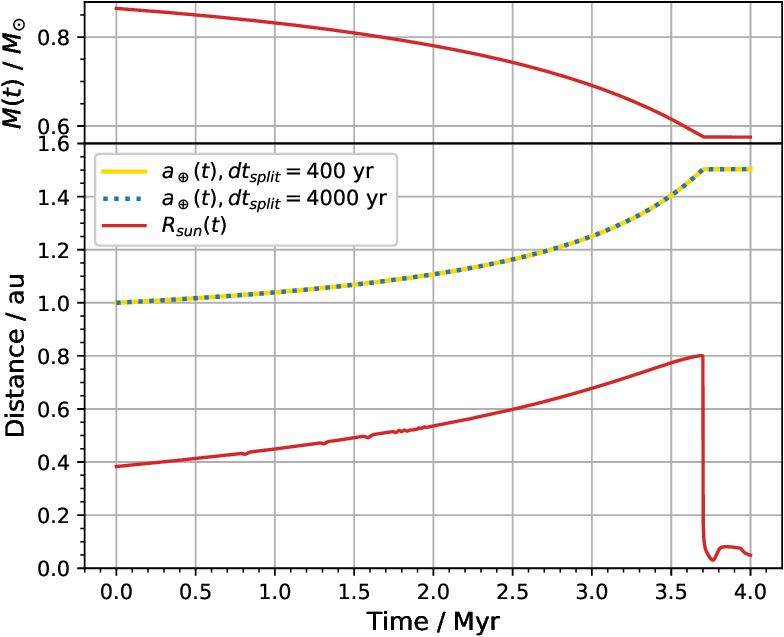}
 \caption{Evolution of the Sun's mass $M(t)$ and radius $R(t)$ (both in solid red) and Earth's semi-major axis $a(t)$ for splitting intervals \dts\ of 400 yr (solid yellow) and 4000 yr (dotted blue).  The simulation starts approximately 4 Myr before the TRGB phase.  Earth's orbital radius starts at 1~au.}
 \label{fig:expansion}
\end{figure}

Fig.~\ref{fig:expansion} shows the Sun's mass and radius compared with Earth's semi-major axis for the two splitting intervals \dts, as functions of simulation time; note this mass-loss profile corresponds to a narrow 4-Myr window around the TRGB seen in Fig.~\ref{fig:massrad}'s $\eta = 0.8$ track.  The solar radius only reaches 0.8~au.  As we expect, Earth's orbit adiabatically expands in sync with solar mass loss, stopping at about 1.5~au when the Sun reaches its TRGB.  Comparing the semi-major axis plots for the two values for \dts\ shows the solutions are indistinguishable and converged.

\S~\ref{sec:convergence} shows more extensive convergence tests, with additional demonstrations of PI in \S~\ref{sec:survey}.  \S~\ref{sec:time_perf} measures the minimal cost in computational overhead associated with our low-level implementation.  Finally, the \jupyter\ Notebooks for PI's documentation (see footnote 9) or to generate Fig.~\ref{fig:expansion} (see ``Data Availability'' section at the end) feature examples of its simple setup and demonstrate its ease of use.

\section{Tides Constant Time Lag (TCTL)}
\label{sec:tctl}

\subsection{TCTL \rebx\ Implementation}
\label{sec:tctl_impl}

We implement a general form of the weak friction model for tidal interaction in binary systems with constant time lag from \citet{Hut1981} \citep[see also][]{Bolmont2015}.  The tidal perturbing force from \citet[Eq. 8]{Hut1981} is
\begin{equation}
 \bm{F} = -G\dfrac{Mm}{r^2}\left\{\hat{r} + 3q\left(\dfrac{R}{r}\right)^5 k_2\left[\left(1 + 3\dfrac{\dot{r}}{r}\tau\right)\hat{r} - (\Omega - \dot{\theta})\tau\hat{\theta}\right] \right\},
 \label{eq:hut}
\end{equation}
where $G$ is the gravitational constant; $M$ and $m$ are the masses of the tidally deformed body and perturber respectively; $r$ is the radial distance between the two as point masses; $q = m/M$ is the mass ratio; $R$ is the perturbed body's physical radius; $\tau$ is a small constant time lag that corresponds to the slight change in both amplitude and direction (i.e., misalignment) of the tides; $\Omega$ and $\dot \theta$ are the rotational (spin) and instantaneous orbital angular velocities of the perturbed body and perturber respectively ($\theta$ is the true anomaly); and $\hat r$ and $\hat \theta$ are unit vectors in the $r$ and $\theta$ directions.

The perturbed body's tidal Love number, $k_2$, is defined as \citep[e.g.,][]{Becker2013},
\begin{equation}
 k_2 = \frac{3 - \eta_2}{2 + \eta_2},
 \label{eq:love_no}
\end{equation}
where $\eta_2$ is the solution of Radau's equation for $j=2$ at the body's surface.  \citet{Hut1981} confusingly refers to this quantity as the apsidal motion constant $k$, which instead would imply a coefficient of $6$ in the $k_2$ term in Eq.~\ref{eq:hut} \citep[e.g.,][]{Csizmadia2019}.  We therefore follow the more standard notation of \cite{Bolmont2015}.

We release this implementation of `Tides Constant Time Lag' (TCTL) in version 3.0.5 of \rebx.\footnote{Documentation is available at \url{https://reboundx.readthedocs.io/en/latest/effects.html\#tides-constant-time-lag}.}  When activated, the tidal effect applies to all other bodies in a \reb\ simulation, allowing for arbitrary orbital inclinations and eccentricities.  Tides can be raised on either the primary or the orbiting bodies -- or both -- by setting the requisite parameters on all desired particles.  For example, if we set a physical radius for the primary, any orbiting body, with non-zero mass, will raise tides on the primary.  Similarly, if we add a physical radius and $k_2$ to any of the orbiting bodies, the primary will raise tides on those particles, e.g., modeling binary star systems.  We note that for computational efficiency, secondary bodies themselves (i.e., all particles added to the simulation beyond the first) will not raise tides on one another with the current implementation.

The inclusion of a non-zero constant time lag $\tau$ introduces dissipation to the system, whereas $\tau = 0$ corresponds to the case of instantaneous equilibrium tides.  The latter case provides a conservative, radial, non-Keplerian potential, i.e., the total energy will be conserved, but the pericentre will precess.  However, in the former case a delayed response typically causes eccentricity damping and will drive orbiting bodies radially either inward or outward depending on whether they orbit faster or slower than the spin ($\Omega$) of the tidally deformed body.

There are two main limitations with the current implementation.  First, the effect does not evolve the spins; it is thus applicable to cases where the angular momentum change due to tides has a negligible effect on the spins or in cases where $\dot{\theta} \ll \Omega$.  Thus, users must consider whether more angular momentum is being exchanged in the system than is available in the spins.  Second, it assumes all of the bodies' spins remain fixed along the reference $z$-axis.  Thus if a body's orbit is inclined with respect to the $xy$-plane, then its spin will be inclined with respect to its orbital plane.

\subsection{TCTL Demonstration}
\label{sec:tctl_demo}

We compare the results of our code with an analytic approximation of Earth's orbital decay around a non-rotating RGB Sun.  We use the following tidal evolution equation derived in \citet[Eq. 9]{Hut1981} to predict the decay of Earth's orbit as a function of time:
\begin{equation}
 \begin{aligned}
 \dfrac{da}{dt} =& -6\dfrac{k_2}{T}q(1 + q)\left(\dfrac{R}{a}\right)^8\dfrac{a}{(1-e^2)^{15/2}}\\
 &\cdot\left\{f_1(e^2) - (1-e^2)^{3/2}f_2(e^2)\dfrac{\Omega}{n}\right\},
 \label{eq:dadt}
 \end{aligned}
\end{equation}
where
\begin{align*}
 f_1(e^2) &= 1 + \tfrac{31}{2}e^2 + \tfrac{255}{8}e^4 + \tfrac{185}{16}e^6 + \tfrac{25}{64}e^8,\\
 f_2(e^2) &= 1 + \tfrac{15}{2}e^2 + \tfrac{45}{8}e^4 + \tfrac{5}{16}e^6,
 \label{eq:fs}
\end{align*}
$n = G^{1/2}(M+m)^{1/2}a^{-3/2}$ is the mean orbital angular velocity, and
\begin{equation*}
 T = \dfrac{R^3}{GM\tau}
\end{equation*}
`is a typical time scale on which significant changes in the orbit take place through tidal evolution' \citep[p. 128]{Hut1981}.  We assume a circular orbit ($e = 0$) and solve differential Eq.~\ref{eq:dadt} to get a predictive expression for Earth's semi-major axis as a function of time:
\begin{equation}
 a(t) = R\left[\left(\dfrac{a_0}{R}\right)^8 - 48\dfrac{k_2}{T}q(1+q)t\right]^{1/8}.
 \label{eq:a}
\end{equation}

We set up an idealised Sun-Earth system just before the TRGB, with $M = 0.86~M_{\sun}$, $R = 0.85~$au, and $\Omega = 0$.  All solar parameters remain constant, and Earth's initial semi-major axis is at 1~au.  We vary Earth's initial eccentricity in two different setups, with $e_{\earth} = 0$ and $e_{\earth} = 0.03$.  Meanwhile, $k_2$ is approximately equal to $\lambda_2$ \citep{Zahn1989, Zahn1977}, which depends on properties of the Sun's convective envelope.  Following \citet[p. 6]{Schroder2008}, we set $k_2 \approx \lambda_2 \approx 0.03$ to be constant for a fully convective envelope.

For a highly-evolved RGB Sun, tidal friction in the outer convective envelope will retard tidal bulges on the solar photosphere \citep{Schroder2008}, resulting in a non-zero value for $\tau$.  Setting Eq. 11 in \citet{Zahn1989} equal to the azimuthal ($\hat{\theta}$) component of our Eq.~\ref{eq:hut} and solving for $\tau$, we find
\begin{equation}
 \tau = \dfrac{2R^3}{GMt_f},
 \label{eq:tau}
\end{equation}
where $t_f(t) = (M(t)R(t)^2/L(t))^{1/3} \approx \mathcal{O}(1 \textrm{yr}$) is the convective friction time \citep[Eq. 7]{Zahn1989}.
Thus, with $t_f = 1$ yr, and $G = 4\pi^2\,\textrm{au}^3\cdot\textrm{yr}^{-2}\cdot M_{\sun}^{-1}$, we set the constant time lag $\tau = 0.04$ yr in both setups.\footnote{A \jupyter\ Notebook containing these tidal tests can be found at \url{https://github.com/dtamayo/reboundx/blob/master/ipython_examples/TidesConstantTimeLag.ipynb}.}

\begin{figure}
 \includegraphics[width=\columnwidth]{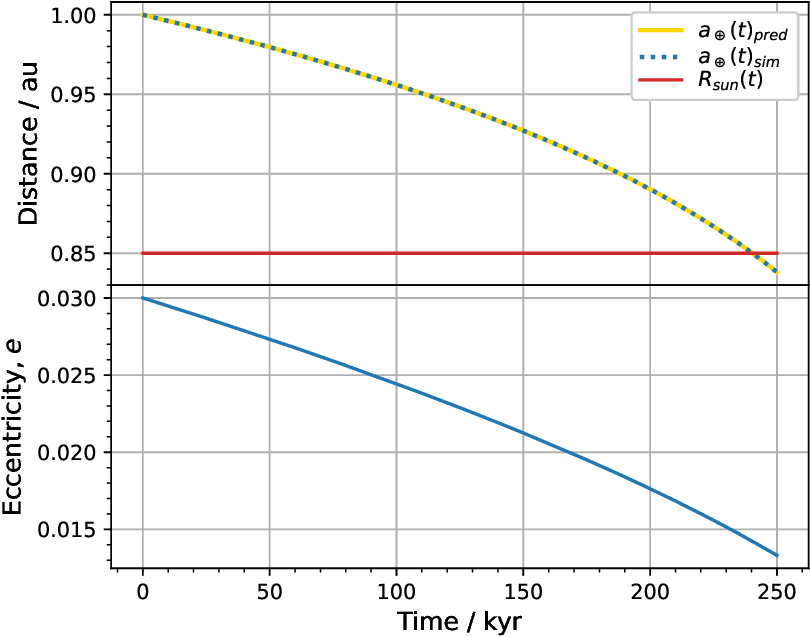}
 \caption{An idealized 250-kyr simulation of the Earth's orbital decay and engulfment due to dissipative tidal interactions with the Sun.  (Top) $a(t)_\textrm{pred}$ and $a(t)_\textrm{sim}$ respectively are the analytically predicted (solid yellow) and simulated (dotted blue) evolutions of Earth's semi-major axis; cf. $R(t)$, the solar radius (red).  (Bottom) A similar setup where Earth's orbital eccentricity (solid blue), initialised to $e_{\earth} = 0.03$, dampens over time due to dissipative tides.}
 \label{fig:tides}
\end{figure}

We plot results for a 250-kyr integration in Fig.~\ref{fig:tides}.  In the top subplot, we see the dissipative tidal effect causes Earth's orbit, measured by its semi-major axis (dotted blue), to decay into the solar photosphere (dashed red).  We run the simulation with \ias\ to best compare our results, $a(t)_\textrm{sim}$, with the theoretical decay, $a(t)_\textrm{pred}$, predicted by Eq.~\ref{eq:a} (solid yellow).  As we can see, the two lines are indistinguishable.  Thus, our numerical results match analytic predictions, validating our low-level implementation.  In the bottom subplot, in our variation with an initial $e_{\earth} = 0.03$, we observe eccentricity damping due to the dissipative tidal effect, consistent with physical expectations.

\section{Combining Effects}
\label{sec:combined}

To further showcase the capabilities of these new features in \rebx, we demonstrate both dynamical stellar evolution (via \S~\ref{sec:ss}) and the effects of dissipative tidal interactions (via \S~\ref{sec:tctl}) running simultaneously.  We use Eq. \ref{eq:tau}, which is solely in terms of stellar mass, radius, and luminosity, as all these values are generated from \mesa.  We then interpolate the time-varying solar data, generated in \S~\ref{sec:pi_demo}, to evaluate and update the corresponding TCTL parameter $\tau$ (\S~\ref{sec:tctl_impl}) throughout a simulation.

One can interpolate stellar mass, radius, and luminosity data separately to evaluate and update $\tau$ with Eq.~\ref{eq:tau} as needed throughout a simulation.  However, as discussed in \S~\ref{sec:convergence} and \S~\ref{sec:time_perf}, the computational overhead associated with excessive interpolation calls can result in increased simulation runtimes.  Since the stellar profiles for $R(t)$, $M(t)$, and $L(t)$ are known in advance from \mesa's output, we instead precalculate the values of $\tau(t)$ with Eq.~\ref{eq:tau} for use with its own interpolator object (\S~\ref{sec:pi_impl}).  This requires only one interpolation call per update of $\tau$ and is therefore more computationally efficient.

The remaining two tidal parameters are $\Omega$ and $k_2$ (see \S~\ref{sec:tctl_impl}).  As conservation of angular momentum and post-MS magnetic braking effectively result in a non-rotating RGB Sun \citep{Schroder2008}, we set $\Omega = 0$ in the following simulations.  Finally, as explained in \S~\ref{sec:tctl_demo}, we set $k_2 = 0.03$ to be constant.

\begin{figure}
 \includegraphics[width=\columnwidth]{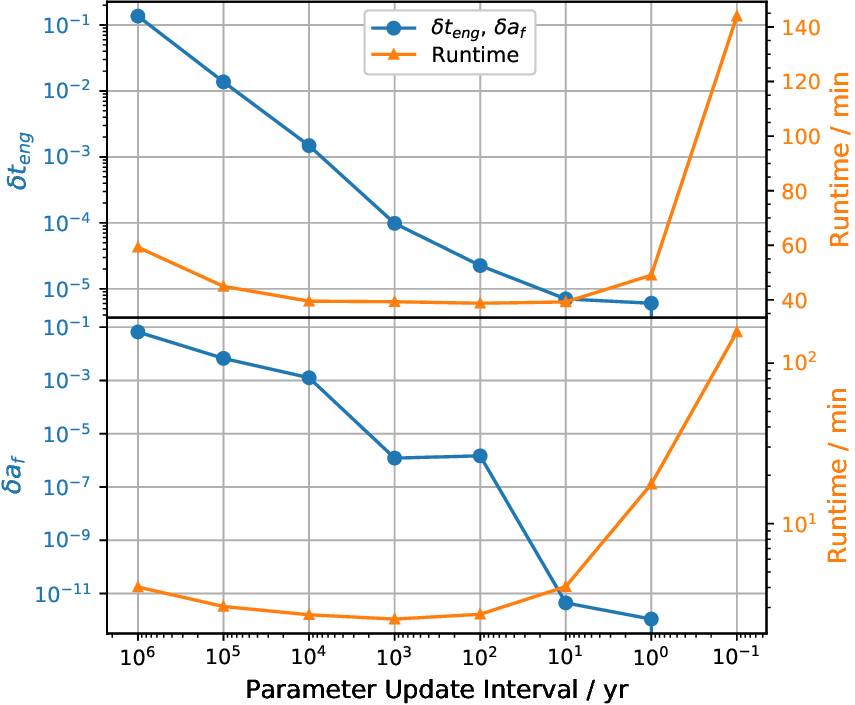}
 \caption{Relative errors of convergence tests of dynamical results as a function of stellar- and tidal-parameter update intervals of two-body, post-MS systems approximately 5 Myr pre-TRGB.  The top subplot shows the relative error (Eq.~\ref{eq:rel_err}) in engulfment times $\delta t_\textrm{eng}$ (blue circles) and simulation runtimes (orange triangles) versus update intervals for an Earth-mass planet with initial semi-major axis of 0.7~au.  The bottom subplot shows the relative error in final semi-major axes $\delta a_\textrm{f}$ (blue circles) and simulation runtimes (orange triangles) versus update intervals for a Jupiter-mass planet with initial semi-major axis of 5~au.}
 \label{fig:converge}
\end{figure}

\begin{figure*}
 \includegraphics[width=\textwidth]{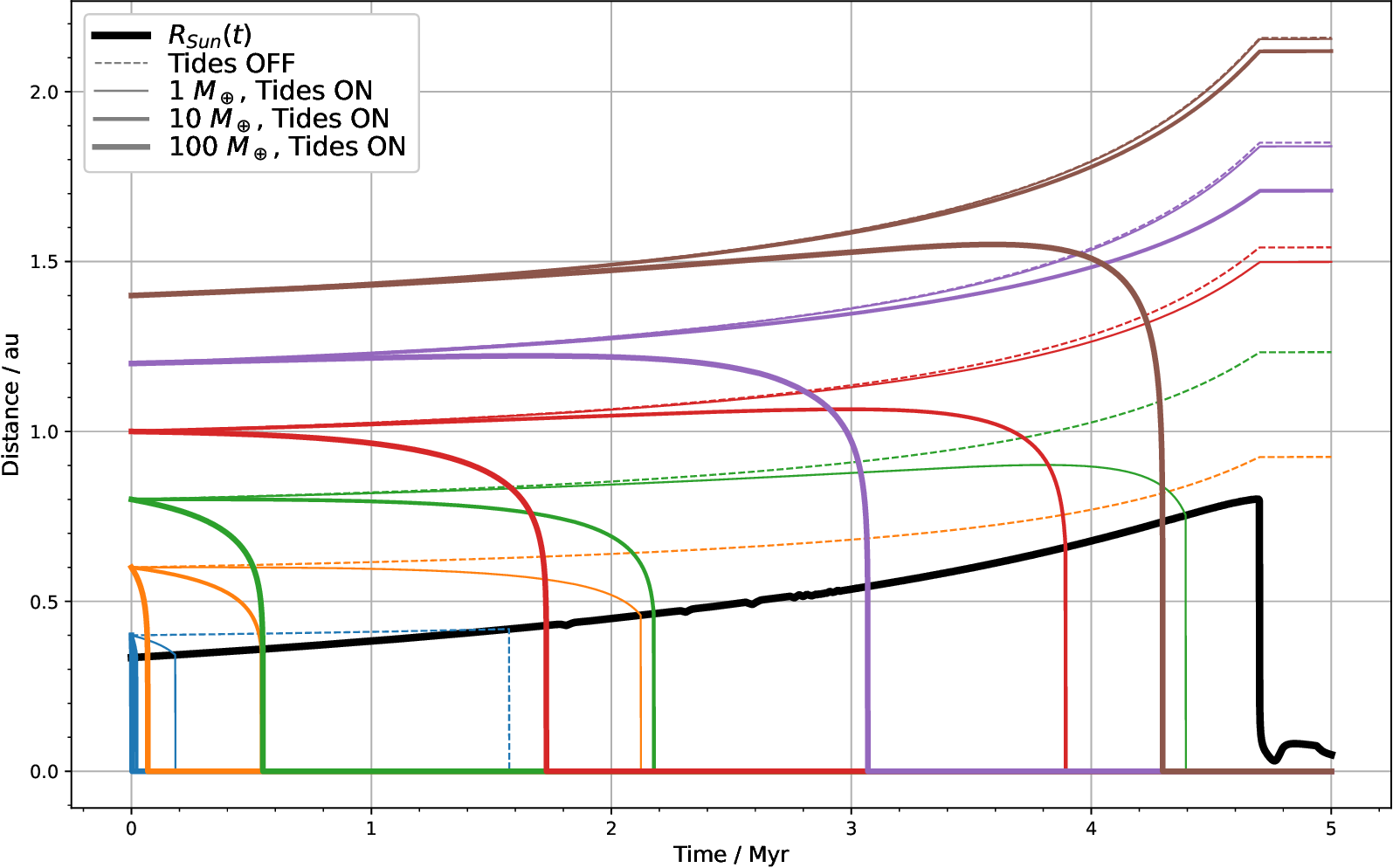}
 \caption{Suites of simulations for the case of a single planet around the Sun, approximately 5 Myr before the TRGB, with TCTL (\S~\ref{sec:tctl}) enabled (solid coloured curves) and disabled (dashed coloured curves), and evolving solar radius (thick black curve) and mass using PI of \mesa\ data (\S~\ref{sec:ss}).  Initial semi-major axes of the planets range from 0.4 to 1.4~au, in increments of 0.2, with each starting location distinguished by its own colour.  The heavier line weights of the solid curves correspond to more massive planets as shown in the legend.  With tides off, planets starting at the same semi-major axis follow the same trajectory regardless of mass.}
 \label{fig:survey}
\end{figure*}

\subsection{Convergence Tests}
\label{sec:convergence}

\citet[\S~14]{Veras2016a} states that combinations of \Nbody\ and stellar evolution codes should ensure errors converge as timesteps decrease.  Thus, in considering \citet[Fig.~1]{Veras2013}, we study the effect that various time intervals between parameter updates has on dynamical results in two different convergence tests.  The first compares the engulfment time of a body closely-orbiting an RGB Sun against a range of time intervals for updating parameters.  The second compares the final semi-major axis reached by a more distant body at the TRGB as a function of the same range of intervals.  We use the high-accuracy \ias\ integrator for all setups and record each runtime.

As seen in Fig.~\ref{fig:survey}, Venus-like planets (i.e., $1~M_{\earth}$ between 0.6 and 0.8~au) cannot escape engulfment by the TRGB when dissipative tidal interactions are considered.  Meanwhile, Earth-like planets at 1~au can survive with or without tides.  Thus, in our first test, we initialise an Earth-mass planet at 0.7~au, about 5 Myr before the TRGB.  We enable both stellar evolution and dissipative tidal interactions, and we record both the integration time when \reb\ detects a particle collision (i.e., the planet is engulfed), $t_\textrm{eng}$, and the elapsed (wall-clock) real time of the simulation.  We interpolate and update the RGB Sun's mass, radius, and time-lag $\tau$ at regular intervals and repeat the runs across a logarithmic range in decades from every 1-Myr to one-tenth a year.

We take our highest accuracy result of the engulfment time, $t_\textrm{eng} = 3.2997761626440026$ Myr for an update interval of 0.1-yr, as our true value.  We then calculate the relative error defined by
\begin{equation}
 \delta t_\textrm{eng} = \frac{|t_\textrm{eng,0} - t_\textrm{eng}|}{t_\textrm{eng}},
 \label{eq:rel_err}
\end{equation}
where $t_\textrm{eng,0}$ is the engulfment time measured at each update interval.  Our results of engulfment-time relative errors versus update intervals can be seen in the top subplot of Fig. \ref{fig:converge}.

We note a difference of almost 0.5 Myr in engulfment time between the least frequent (every Myr) and the most frequent update intervals (ten times per year).  The shortest update intervals (yearly and $10^{-1}$-yr) coincide with noticeable increases in total runtimes.  The case with the most frequent updates takes more than three times longer to run than the fastest simulation with $10^2$-yr updates.  Comparing the two curves, the additional computational overhead from excessive interpolation and updating yields diminishing returns to accuracy.

Looking to the left-hand side of the subplot, between the $10^6$- and $10^4$-yr intervals, we find runtimes first start out longer than those around the middle and decrease with shorter intervals.  Since our runs terminate upon engulfment, this behaviour corresponds to instances where the planet is able to survive longer due to a slower orbital decay.  Rewriting Eq.~\ref{eq:a} for the analytic approximation of the planet's semi-major axis as a function of time, we find
\begin{equation}
 a(t) = \left[a_0^8 - 48R^5GM\tau k_2q(1+q)t\right]^{1/8}.
 \label{eq:decay}
\end{equation}
For a positive time-lag $\tau$, inspection of Eq.~\ref{eq:decay} reveals that an increase in solar radius $R$ results in a decrease in semi-major axis $a(t)$.  Thus shorter parameter update intervals that more accurately capture the rapid growth of the TRGB solar radius serve to accelerate orbital decay toward engulfment.  In other words, until the $10^2$- and $10$-yr range, more frequent updates result in shorter runtimes since engulfment occurs sooner.  Conversely, longer update intervals capture radial growth less accurately, helping to slow orbital decay and resulting in the longer aforementioned engulfment times.

In our second test, we initialise a Jupiter-mass at 5~au at the same solar age.  Stellar evolution and dissipative tides remain enabled.  As engulfment by the TRGB does not occur, we record the final semi-major axis of the planet, $a_f$, after a full 5-Myr integration.  We repeat the simulation for the same range of parameter update intervals as before and take $a_f = 7.7047437314161416$~au from our update interval of 0.1-yr as our true value.

Following Eq.~\ref{eq:rel_err}, we plot the relative error of our results in the bottom subplot of Fig. \ref{fig:converge}.  We note a difference of about 0.5~au in final semi-major axis reached by the planet between the longest ($10^6$ yr) and shortest ($10^{-1}$ yr) update intervals.  Again we find that excessive interpolation and updating, between intervals of $10$- and $10^{-1}$-yr, result in longer computational runtimes (more than 60 times in the worst case) with diminishing returns in accuracy.

\subsection{Engulfment Survey}
\label{sec:survey}

Similar to \citet{Rasio1996}, \citet{Rao2018}, \citet{Villaver}, and \citet[\S~5]{Veras2016a}, we examine the effects of stellar mass-loss, dissipative tides, and planet mass on the orbital evolution of close-in planets around an RGB Sun.

We survey several suites of single-planet setups about 5 Myr before the TRGB.  We include stellar evolution in all cases and run each setup twice: once with TCTL on and once with it off.  We use the \ias\ integrator, and we opt for a parameter update interval of every 100 yr based on Fig. \ref{fig:converge}'s results in \S~\ref{sec:convergence}.  Our three main testing suites involve a single planet of either 1, 10, or 100 Earth-masses (the latter two are comparable to the masses of Uranus and Saturn, respectively).  For each suite we initialise the planet's semi-major axis between 0.4 and 1.4~au in increments of 0.2.  We choose a lower bound for the orbital distance of 0.4~au because the solar radius is already larger than 0.3~au at the start of the 5 Myr integrations.

We show the results of our survey in Fig. \ref{fig:survey}.  The thick black curves correspond to the RGB Sun's radius, reaching its tip around 4.7 Myr into the simulation (cf. Figs. \ref{fig:massrad} and \ref{fig:expansion}).  The solid and dashed coloured curves correspond to the planet's semi-major axis with and without tides present, respectively.

We first note that the planet's orbit in non-tidal cases (dashed coloured curves) all exhibit the same adiabatic expansion due to the stellar mass loss, stopping once the Sun reaches the TRGB (cf. \S~\ref{sec:pi_demo} and Fig. \ref{fig:expansion}).  Differences in the final semi-major axis reached without tides depend only on initial semi-major axis with no dependence on planetary mass.  Among these non-tidal cases, engulfment occurs only for planets with an initial semi-major axis of 0.4~au (dashed blue curve).  This is consistent with the non-tidal results of \citet[Fig.~7]{Sackmann1993}, namely Mercury's engulfment and the survival of outer terrestrial planets by the TRGB.

With TCTL enabled (solid coloured curves), we observe the tidal drag effect begin to dominate adiabatic expansion (see \S~\ref{sec:tctl}).  In the 1-$M_{\earth}$ suite (thinnest solid curves), we see that drag on the planet from tides raised on the Sun result in engulfment by the TRGB for initial semi-major axes between 0.4 and 0.8~au; this is consistent with the results of \citet[Fig.~5]{Rao2018} and, in particular, Venus' engulfment and Earth's survival in \citet[Fig.~2]{Rasio1996}.  We find similar results between $0.4 \leq a_0 \leq 1.0$~au for the 10-$M_{\earth}$ (Uranus-like) suite (thicker solid curves), and between $0.4 \leq a_0 \leq 1.4$~au for the 100-$M_{\earth}$ (Saturn-like) suite (thickest solid curves).

The wider range of $a_0$ that lead to engulfment as a function of planetary mass is mathematically consistent with the tidal force being directly proportional to the perturbing mass ($m$ in Eq. \ref{eq:hut}) and physically consistent with raising larger tidal bulges on the Sun's surface which lag behind the planet's orbit.  This agrees with findings in \citet{Villaver}: (1) planet engulfment along the red giant branch is quite sensitive to the planetary mass; and (2) the range of initial separations for planet engulfment increases with planetary mass.  As a final note, we observe attenuation of adiabatic expansion due to tides in the surviving planetary cases, e.g., $a_0 \geq 1.0$~au for 1 $M_{\earth}$ and $a_0 \geq 1.1$~au for 10 $M_{\earth}$.

\subsection{Time Performance}
\label{sec:time_perf}

To measure the computational performance costs of these two new features,  we record runtimes over multiple trials of the terrestrial planets simultaneously orbiting a pre-TRGB Sun.  The four configurations we specify include (1) no new effects; (2) `Parameter Interpolation' Stellar Evolution (PISE) only; (3) TCTL only; and (4) both effects running simultaneously.  For these runs, we instead use the \whfast\ integrator with a fixed timestep of one-tenth Mercury's initial orbital period to rule out any differences in performance between the four setups caused by adaptive timesteps (e.g., \ias).

We end the integration after 920 kyr for all configurations, which corresponds to the engulfment of Mercury when both stellar evolution and tidal interactions are enabled.  In configurations (2) and (4), we interpolate and update stellar mass, radius and time lag $\tau$ parameters 1000 times throughout the run, corresponding to a frequency interval of 920 yr.  For configuration (3), we evaluate and set $\tau$ only once before the start of the integration.  We perform ten single-threaded runs of each setup on a computing cluster with each node containing two Intel Xeon E5-2640v3 (8-core) CPUs and 128 GB of available memory.

\begin{table}
 \caption{Computational time performance results from 920 kyr simulations of all four terrestrial planets in various \rebx\ configurations, using the WHFast integrator with fixed timesteps.  We computed the average and standard deviation of 10 runs for each of the following setups: no \rebx\ effects (None); `Parameter Interpolation' Stellar Evolution only (PISE); tidal interaction only through TCTL; and both effects simultaneously (PISE \& TCTL).}
 \label{tab:performance}
 \centering
 \begin{tabular}{lcccc}
  \hline
  \multicolumn{1}{c}{\textrm{Effects}} & \textrm{Avg. Runtime} & \textrm{Std. Dev.} & \textrm{Increase}\\
   & \textrm{(s)} & \textrm{(s)} & \textrm{(\%)}\\
  \hline
  None & 57.73 & $\pm 0.37$ &\\
  PISE & 58.69 & $\pm 0.37$ & +1.7\\
  TCTL & 67.30 & $\pm 0.74$ & +16.6\\
  PISE \& TCTL & 68.26 & $ \pm 0.62$ & +18.2\\
  \hline
 \end{tabular}
\end{table}

Table~\ref{tab:performance} shows the computed averages, standard deviations, and percentage increases of runtimes for each configuration.  We find including PISE alone adds (on average) less than a 2 per cent increase in overhead.  The addition of TCTL alone adds an average of 17 per cent to the computation time.  As expected from the above benchmarks, including both effects extends the runtime by about 18 per cent.  While exact runtimes will vary depending on hardware, these increases in overhead are not prohibitive for extended integrations, e.g., on the order of hundreds of millions or billions of orbits.

\subsection{Giant Planets Expansion}
\label{sec:giants}

Similar to \citet[]{Veras2016b}, but without an additional distant planet or the effect of Galactic tides, we simulate the Sun's post-MS influence on the outer giant planets using the $\eta = 0.5$ evolutionary track (Fig.~\ref{fig:massrad}).  We initiate Jupiter, Saturn, Uranus, and Neptune using \reb's built-in ability to add particles by obtaining ephemerides from NASA's HORIZONS database.\footnote{See  \url{https://rebound.readthedocs.io/en/latest/ipython_examples/Horizons/} for more information.}  The 250-Myr simulation and PISE of \mesa\ data begin about 110 Myr before the TRGB with a fixed \whfast\ timestep of 0.5 yr.  As the Sun's mass is about 0.993 $M_{\sun}$ by the start of this age, we use the PI's cubic spline to allow a 5-Myr smooth, sigmoid transition from 1 $M_{\sun}$ to avoid any instabilities upfront.

\begin{figure}
 \includegraphics[width=\columnwidth]{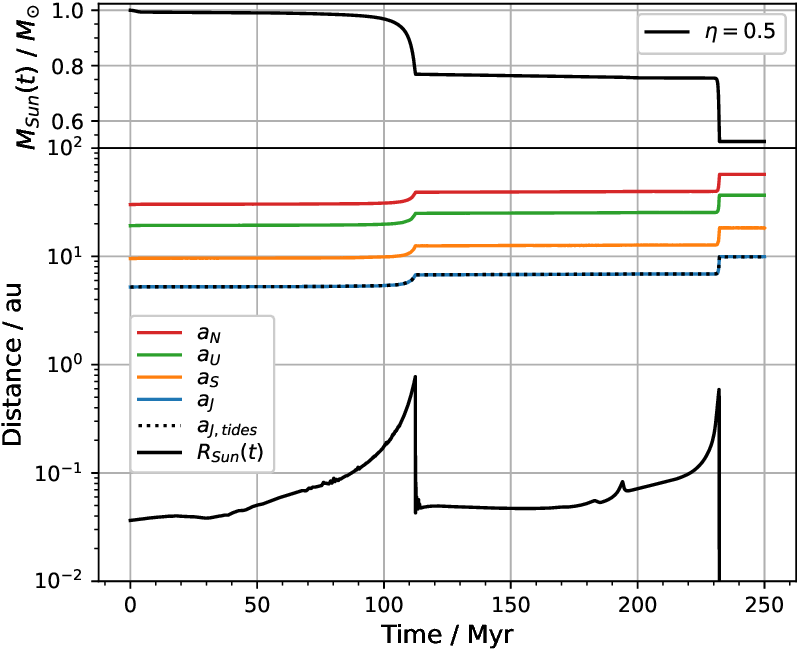}
 \caption{Post-MS evolution of the Sun and the Solar System's outer giant planets throughout the RGB and AGB phases until the start of the WD phase.  The top panel shows solar mass-loss (solid black) for an $\eta = 0.5$ scaling factor (Eq.~\ref{eq:reimers}; cf. Fig.~\ref{fig:massrad}) but with a 5 Myr sigmoid initial transition from 1 $M_{\sun}$ to $~0.993\,M_{\sun}$.  The bottom panel shows the Sun's radial extent (solid black) and the semi-major axes of Jupiter, Saturn, Uranus, and Neptune (solid colours); we also include Jupiter's semi-major axis in a TCTL enabled run (dotted black).}
 \label{fig:giants}
\end{figure}

Fig.~\ref{fig:giants} shows adiabatic expansion of the giant planets' orbits, corresponding with solar mass-loss, until the WD phase, about 230 Myr in.  For clarity, we plot only Jupiter's semi-major axis in our TCTL-enabled run, as its effect is imperceptible at these scales.  Our results are very consistent with Fig.~4 of \citet{Veras2016b}; e.g., Jupiter and Saturn's final semi-major axes roughly reach Saturn and Uranus' initial semi-major axes, respectively.  Finally, the TCTL-free and TCTL runs completed in just under 8 and 9 minutes, respectively, on a modern laptop with an Intel Core i5-8265U CPU.

\section{Conclusion}
\label{sec:Conclusion}

We add two new features to \rebx's existing library of astrophysical effects: generalised parameter interpolation for splitting schemes (\S~\ref{sec:ss}) and dissipative tidal interactions (\S~\ref{sec:tctl}).  The former conveniently allows the results of other integration codes to be used as parameter inputs for \reb.  The latter lets users examine tidal effects among close encounter situations, e.g., `hot Jupiters' around post-MS stars.  Users can also utilise both features simultaneously (\S~\ref{sec:combined}) to study in detail a wide-range of orbital instabilities caused by stellar mass loss and tidal drag, common subjects in the active post-MS planetary science field.

The main advantages of our implementations are their usability with \reb's various integrators and interoperability with other codes.  We show these two new features to be consistent with prior works as well as accurate and efficient by our convergence and performance studies.  We hope that these improvements will facilitate new numerical investigations and encourage others in the community to continue contributing to the \rebx\ library.

\section*{Acknowledgements}

We thank Tam\'as Borkovits and Ricardo Moraes for helpful discussions.
Simulations in this paper made use of the \mesa, \reb\ and \rebx\ codes, all of which are freely available at \url{http://mesa.sourceforge.net/}, \url{http://github.com/hannorein/rebound}, and \url{https://github.com/dtamayo/reboundx}.  This research was made possible by the open-source projects \jupyter\ \citep{Jupyter}, \texttt{IPython}
\citep{IPython}, and \texttt{matplotlib} \citep{matplotlib1, matplotlib2}.

The \mesa\ EOS is a blend of the OPAL \citet{Rogers2002}, SCVH
\citet{Saumon1995}, PTEH \citet{Pols1995}, HELM
\citet{Timmes2000}, and PC \citet{Potekhin2010} EOSes.  Radiative opacities are primarily from OPAL \citep{Iglesias1993,
Iglesias1996}, with low-temperature data from \citet{Ferguson2005}
and the high-temperature, Compton-scattering dominated regime by
\citet{Buchler1976}.  Electron conduction opacities are from
\citet{Cassisi2007}.  Nuclear reaction rates are a combination of rates from
NACRE \citep{Angulo1999}, JINA REACLIB \citep{Cyburt2010}, plus
additional tabulated weak reaction rates \citet{Fuller1985, Oda1994,
Langanke2000}.  (For \mesa\ versions before 11701): Screening is
included via the prescriptions of \citet{Salpeter1954, Dewitt1973,
Alastuey1978, Itoh1979}.  (For \mesa\ versions 11701 or later): Screening
is included via the prescription of \citet{Chugunov2007}.  Thermal
neutrino loss rates are from \citet{Itoh1996}.

All simulations performed in \S~\ref{sec:combined} were run on the `Penguin' Cherry-Creek 2 cluster at the UNLV National Supercomputing Institute for High Performance Computing and Communications in Nevada (\url{https://www.nscee.edu/}).  Last, but not least, the authors thank the referee for constructive comments that improved this manuscript.

\section*{Data availability}

The data underlying this article, including \python\ scripts to generate all the figures, are available on GitHub, at \url{https://github.com/sabaronett/REBOUNDxPaper}.

%%%%%%%%%%%%%%%%%%%%%%%%%%%%%%%%%%%%%%%%%%%%%%%%%%



\bibliographystyle{mnras}
\bibliography{references} % if your bibtex file is called example.bib

\begin{thebibliography}{}
\makeatletter
\relax
\def\mn@urlcharsother{\let\do\@makeother \do\$\do\&\do\#\do\^\do\_\do\%\do\~}
\def\mn@doi{\begingroup\mn@urlcharsother \@ifnextchar [ {\mn@doi@}
  {\mn@doi@[]}}
\def\mn@doi@[#1]#2{\def\@tempa{#1}\ifx\@tempa\@empty \href
  {http://dx.doi.org/#2} {doi:#2}\else \href {http://dx.doi.org/#2} {#1}\fi
  \endgroup}
\def\mn@eprint#1#2{\mn@eprint@#1:#2::\@nil}
\def\mn@eprint@arXiv#1{\href {http://arxiv.org/abs/#1} {{\tt arXiv:#1}}}
\def\mn@eprint@dblp#1{\href {http://dblp.uni-trier.de/rec/bibtex/#1.xml}
  {dblp:#1}}
\def\mn@eprint@#1:#2:#3:#4\@nil{\def\@tempa {#1}\def\@tempb {#2}\def\@tempc
  {#3}\ifx \@tempc \@empty \let \@tempc \@tempb \let \@tempb \@tempa \fi \ifx
  \@tempb \@empty \def\@tempb {arXiv}\fi \@ifundefined
  {mn@eprint@\@tempb}{\@tempb:\@tempc}{\expandafter \expandafter \csname
  mn@eprint@\@tempb\endcsname \expandafter{\@tempc}}}

\bibitem[\protect\citeauthoryear{{Alastuey} \& {Jancovici}}{{Alastuey} \&
  {Jancovici}}{1978}]{Alastuey1978}
{Alastuey} A.,  {Jancovici} B.,  1978, \mn@doi [\apj] {10.1086/156681}, \href
  {https://ui.adsabs.harvard.edu/\#abs/1978ApJ...226.1034A} {226, 1034}

\bibitem[\protect\citeauthoryear{{Angulo} et~al.,}{{Angulo}
  et~al.}{1999}]{Angulo1999}
{Angulo} C.,  et~al., 1999, \mn@doi [Nuclear Physics A]
  {10.1016/S0375-9474(99)00030-5}, \href
  {https://ui.adsabs.harvard.edu/abs/1999NuPhA.656....3A} {656, 3}

\bibitem[\protect\citeauthoryear{{Becker} \& {Batygin}}{{Becker} \&
  {Batygin}}{2013}]{Becker2013}
{Becker} J.~C.,  {Batygin} K.,  2013, \mn@doi [\apj]
  {10.1088/0004-637X/778/2/100}, \href
  {https://ui.adsabs.harvard.edu/abs/2013ApJ...778..100B} {778, 100}

\bibitem[\protect\citeauthoryear{{Blöcker}}{{Blöcker}}{1995}]{Blocker1995}
{Blöcker} T.,  1995, \aap, \href
  {https://ui.adsabs.harvard.edu/abs/1995A&A...297..727B} {297, 727}

\bibitem[\protect\citeauthoryear{{Bolmont}, {Raymond}, {Leconte}, {Hersant}  \&
  {Correia}}{{Bolmont} et~al.}{2015}]{Bolmont2015}
{Bolmont} E.,  {Raymond} S.~N.,  {Leconte} J.,  {Hersant} F.,   {Correia} A.
  C.~M.,  2015, \mn@doi [\aap] {10.1051/0004-6361/201525909}, \href
  {https://ui.adsabs.harvard.edu/abs/2015A&A...583A.116B} {583, A116}

\bibitem[\protect\citeauthoryear{{Buchler} \& {Yueh}}{{Buchler} \&
  {Yueh}}{1976}]{Buchler1976}
{Buchler} J.~R.,  {Yueh} W.~R.,  1976, \mn@doi [\apj] {10.1086/154847}, \href
  {https://ui.adsabs.harvard.edu/abs/1976ApJ...210..440B} {210, 440}

\bibitem[\protect\citeauthoryear{{Cassisi}, {Potekhin}, {Pietrinferni},
  {Catelan}  \& {Salaris}}{{Cassisi} et~al.}{2007}]{Cassisi2007}
{Cassisi} S.,  {Potekhin} A.~Y.,  {Pietrinferni} A.,  {Catelan} M.,   {Salaris}
  M.,  2007, \mn@doi [\apj] {10.1086/516819}, \href
  {https://ui.adsabs.harvard.edu/abs/2007ApJ...661.1094C} {661, 1094}

\bibitem[\protect\citeauthoryear{Caswell et~al.,}{Caswell
  et~al.}{2020}]{matplotlib2}
Caswell T.~A.,  et~al., 2020, matplotlib/matplotlib: REL: v3.3.1,
  \mn@doi{10.5281/zenodo.3984190}, \url
  {https://doi.org/10.5281/zenodo.3984190}

\bibitem[\protect\citeauthoryear{{Chambers}}{{Chambers}}{1999}]{MERCURY}
{Chambers} J.~E.,  1999, \mn@doi [\mnras] {10.1046/j.1365-8711.1999.02379.x},
  \href {https://ui.adsabs.harvard.edu/abs/1999MNRAS.304..793C} {304, 793}

\bibitem[\protect\citeauthoryear{Choi, Dotter, Conroy, Cantiello, Paxton  \&
  Johnson}{Choi et~al.}{2016}]{Choi2016}
Choi J.,  Dotter A.,  Conroy C.,  Cantiello M.,  Paxton B.,   Johnson B.~D.,
  2016, \mn@doi [The Astrophysical Journal] {10.3847/0004-637x/823/2/102}, 823,
  102

\bibitem[\protect\citeauthoryear{{Chugunov}, {Dewitt}  \&
  {Yakovlev}}{{Chugunov} et~al.}{2007}]{Chugunov2007}
{Chugunov} A.~I.,  {Dewitt} H.~E.,   {Yakovlev} D.~G.,  2007, \mn@doi [\prd]
  {10.1103/PhysRevD.76.025028}, \href
  {https://ui.adsabs.harvard.edu/abs/2007PhRvD..76b5028C} {76, 025028}

\bibitem[\protect\citeauthoryear{{Csizmadia}, {Hellard}  \&
  {Smith}}{{Csizmadia} et~al.}{2019}]{Csizmadia2019}
{Csizmadia} S.,  {Hellard} H.,   {Smith} A.~M.~S.,  2019, \mn@doi [\aap]
  {10.1051/0004-6361/201834376}, \href
  {https://ui.adsabs.harvard.edu/abs/2019A&A...623A..45C} {623, A45}

\bibitem[\protect\citeauthoryear{{Cyburt} et~al.,}{{Cyburt}
  et~al.}{2010}]{Cyburt2010}
{Cyburt} R.~H.,  et~al., 2010, \mn@doi [\apjs] {10.1088/0067-0049/189/1/240},
  \href {https://ui.adsabs.harvard.edu/abs/2010ApJS..189..240C} {189, 240}

\bibitem[\protect\citeauthoryear{{Dewitt}, {Graboske}  \& {Cooper}}{{Dewitt}
  et~al.}{1973}]{Dewitt1973}
{Dewitt} H.~E.,  {Graboske} H.~C.,   {Cooper} M.~S.,  1973, \mn@doi [\apj]
  {10.1086/152061}, \href
  {https://ui.adsabs.harvard.edu/\#abs/1973ApJ...181..439D} {181, 439}

\bibitem[\protect\citeauthoryear{{Eggleton}}{{Eggleton}}{1971}]{Eggleton1971}
{Eggleton} P.~P.,  1971, \mn@doi [\mnras] {10.1093/mnras/151.3.351}, \href
  {https://ui.adsabs.harvard.edu/abs/1971MNRAS.151..351E} {151, 351}

\bibitem[\protect\citeauthoryear{{Eggleton}}{{Eggleton}}{1972}]{Eggleton1972}
{Eggleton} P.~P.,  1972, \mn@doi [\mnras] {10.1093/mnras/156.3.361}, \href
  {https://ui.adsabs.harvard.edu/abs/1972MNRAS.156..361E} {156, 361}

\bibitem[\protect\citeauthoryear{{Eggleton}}{{Eggleton}}{1973}]{Eggleton1973}
{Eggleton} P.~P.,  1973, \mn@doi [\mnras] {10.1093/mnras/163.3.279}, \href
  {https://ui.adsabs.harvard.edu/abs/1973MNRAS.163..279E} {163, 279}

\bibitem[\protect\citeauthoryear{{Ferguson}, {Alexander}, {Allard}, {Barman},
  {Bodnarik}, {Hauschildt}, {Heffner-Wong}  \& {Tamanai}}{{Ferguson}
  et~al.}{2005}]{Ferguson2005}
{Ferguson} J.~W.,  {Alexander} D.~R.,  {Allard} F.,  {Barman} T.,  {Bodnarik}
  J.~G.,  {Hauschildt} P.~H.,  {Heffner-Wong} A.,   {Tamanai} A.,  2005,
  \mn@doi [\apj] {10.1086/428642}, \href
  {https://ui.adsabs.harvard.edu/abs/2005ApJ...623..585F} {623, 585}

\bibitem[\protect\citeauthoryear{{Fuller}, {Fowler}  \& {Newman}}{{Fuller}
  et~al.}{1985}]{Fuller1985}
{Fuller} G.~M.,  {Fowler} W.~A.,   {Newman} M.~J.,  1985, \mn@doi [\apj]
  {10.1086/163208}, \href
  {https://ui.adsabs.harvard.edu/abs/1985ApJ...293....1F} {293, 1}

\bibitem[\protect\citeauthoryear{Hairer, Lubich  \& Wanner}{Hairer
  et~al.}{2006}]{Hairer2006}
Hairer E.,  Lubich C.,   Wanner G.,  2006, Geometric numerical integration:
  structure-preserving algorithms for ordinary differential equations.
Springer Science \& Business Media

\bibitem[\protect\citeauthoryear{{Hunter}}{{Hunter}}{2007}]{matplotlib1}
{Hunter} J.~D.,  2007, Computing in Science Engineering, 9, 90

\bibitem[\protect\citeauthoryear{{Hut}}{{Hut}}{1981}]{Hut1981}
{Hut} P.,  1981, \aap, \href
  {https://ui.adsabs.harvard.edu/abs/1981A&A....99..126H} {99, 126}

\bibitem[\protect\citeauthoryear{{Iglesias} \& {Rogers}}{{Iglesias} \&
  {Rogers}}{1993}]{Iglesias1993}
{Iglesias} C.~A.,  {Rogers} F.~J.,  1993, \mn@doi [\apj] {10.1086/172958},
  \href {https://ui.adsabs.harvard.edu/abs/1993ApJ...412..752I} {412, 752}

\bibitem[\protect\citeauthoryear{{Iglesias} \& {Rogers}}{{Iglesias} \&
  {Rogers}}{1996}]{Iglesias1996}
{Iglesias} C.~A.,  {Rogers} F.~J.,  1996, \mn@doi [\apj] {10.1086/177381},
  \href {https://ui.adsabs.harvard.edu/abs/1996ApJ...464..943I} {464, 943}

\bibitem[\protect\citeauthoryear{{Itoh}, {Totsuji}, {Ichimaru}  \&
  {Dewitt}}{{Itoh} et~al.}{1979}]{Itoh1979}
{Itoh} N.,  {Totsuji} H.,  {Ichimaru} S.,   {Dewitt} H.~E.,  1979, \mn@doi
  [\apj] {10.1086/157590}, \href
  {https://ui.adsabs.harvard.edu/\#abs/1979ApJ...234.1079I} {234, 1079}

\bibitem[\protect\citeauthoryear{{Itoh}, {Hayashi}, {Nishikawa}  \&
  {Kohyama}}{{Itoh} et~al.}{1996}]{Itoh1996}
{Itoh} N.,  {Hayashi} H.,  {Nishikawa} A.,   {Kohyama} Y.,  1996, \mn@doi
  [\apjs] {10.1086/192264}, \href
  {https://ui.adsabs.harvard.edu/abs/1996ApJS..102..411I} {102, 411}

\bibitem[\protect\citeauthoryear{Kluyver et~al.,}{Kluyver
  et~al.}{2016}]{Jupyter}
Kluyver T.,  et~al., 2016, in Loizides F.,  Scmidt B.,  eds, Positioning and
  Power in Academic Publishing: Players, Agents and Agendas. IOS Press, pp
  87--90, \url {https://eprints.soton.ac.uk/403913/}

\bibitem[\protect\citeauthoryear{{Langanke} \&
  {Mart{\'{\i}}nez-Pinedo}}{{Langanke} \&
  {Mart{\'{\i}}nez-Pinedo}}{2000}]{Langanke2000}
{Langanke} K.,  {Mart{\'{\i}}nez-Pinedo} G.,  2000, \mn@doi [Nuclear Physics A]
  {10.1016/S0375-9474(00)00131-7}, \href
  {https://ui.adsabs.harvard.edu/abs/2000NuPhA.673..481L} {673, 481}

\bibitem[\protect\citeauthoryear{{Oda}, {Hino}, {Muto}, {Takahara}  \&
  {Sato}}{{Oda} et~al.}{1994}]{Oda1994}
{Oda} T.,  {Hino} M.,  {Muto} K.,  {Takahara} M.,   {Sato} K.,  1994, \mn@doi
  [Atomic Data and Nuclear Data Tables] {10.1006/adnd.1994.1007}, \href
  {https://ui.adsabs.harvard.edu/abs/1994ADNDT..56..231O} {56, 231}

\bibitem[\protect\citeauthoryear{{Paxton}, {Bildsten}, {Dotter}, {Herwig},
  {Lesaffre}  \& {Timmes}}{{Paxton} et~al.}{2011}]{Paxton2011}
{Paxton} B.,  {Bildsten} L.,  {Dotter} A.,  {Herwig} F.,  {Lesaffre} P.,
  {Timmes} F.,  2011, \mn@doi [\apjs] {10.1088/0067-0049/192/1/3}, \href
  {https://ui.adsabs.harvard.edu/abs/2011ApJS..192....3P} {192, 3}

\bibitem[\protect\citeauthoryear{{Paxton} et~al.,}{{Paxton}
  et~al.}{2013}]{Paxton2013}
{Paxton} B.,  et~al., 2013, \mn@doi [\apjs] {10.1088/0067-0049/208/1/4}, \href
  {https://ui.adsabs.harvard.edu/abs/2013ApJS..208....4P} {208, 4}

\bibitem[\protect\citeauthoryear{{Paxton} et~al.,}{{Paxton}
  et~al.}{2015}]{Paxton2015}
{Paxton} B.,  et~al., 2015, \mn@doi [\apjs] {10.1088/0067-0049/220/1/15}, \href
  {https://ui.adsabs.harvard.edu/abs/2015ApJS..220...15P} {220, 15}

\bibitem[\protect\citeauthoryear{{Paxton} et~al.,}{{Paxton}
  et~al.}{2018}]{Paxton2018}
{Paxton} B.,  et~al., 2018, \mn@doi [\apjs] {10.3847/1538-4365/aaa5a8}, \href
  {https://ui.adsabs.harvard.edu/abs/2018ApJS..234...34P} {234, 34}

\bibitem[\protect\citeauthoryear{{Paxton} et~al.,}{{Paxton}
  et~al.}{2019}]{Paxton2019}
{Paxton} B.,  et~al., 2019, \mn@doi [\apjs] {10.3847/1538-4365/ab2241}, \href
  {https://ui.adsabs.harvard.edu/abs/2019ApJS..243...10P} {243, 10}

\bibitem[\protect\citeauthoryear{{Perez} \& {Granger}}{{Perez} \&
  {Granger}}{2007}]{IPython}
{Perez} F.,  {Granger} B.~E.,  2007, Computing in Science Engineering, 9, 21

\bibitem[\protect\citeauthoryear{{Pols}, {Tout}, {Eggleton}  \& {Han}}{{Pols}
  et~al.}{1995}]{Pols1995}
{Pols} O.~R.,  {Tout} C.~A.,  {Eggleton} P.~P.,   {Han} Z.,  1995, \mn@doi
  [\mnras] {10.1093/mnras/274.3.964}, \href
  {https://ui.adsabs.harvard.edu/abs/1995MNRAS.274..964P} {274, 964}

\bibitem[\protect\citeauthoryear{Portegies~Zwart}{Portegies~Zwart}{2018}]{Portegies2018}
Portegies~Zwart S.,  2018, Science, 361, 979

\bibitem[\protect\citeauthoryear{Portegies~Zwart \& McMillan}{Portegies~Zwart
  \& McMillan}{2018}]{Portegies2018book}
Portegies~Zwart S.,  McMillan S.,  2018, Astrophysical Recipes; The art of
  AMUSE, by Portegies Zwart, Simon; McMillan, Steve. ISBN: 978-0-7503-1321-6.
  IOP ebooks. Bristol, UK: IOP Publishing, 2018

\bibitem[\protect\citeauthoryear{{Portegies Zwart}, {Pelupessy},
  {Mart{\'\i}nez-Barbosa}, {van Elteren}  \& {McMillan}}{{Portegies Zwart}
  et~al.}{2020}]{Portegies2020}
{Portegies Zwart} S.,  {Pelupessy} I.,  {Mart{\'\i}nez-Barbosa} C.,  {van
  Elteren} A.,   {McMillan} S.,  2020, \mn@doi [Communications in Nonlinear
  Science and Numerical Simulations] {10.1016/j.cnsns.2020.105240}, \href
  {https://ui.adsabs.harvard.edu/abs/2020CNSNS..8505240P} {85, 105240}

\bibitem[\protect\citeauthoryear{{Potekhin} \& {Chabrier}}{{Potekhin} \&
  {Chabrier}}{2010}]{Potekhin2010}
{Potekhin} A.~Y.,  {Chabrier} G.,  2010, \mn@doi [Contributions to Plasma
  Physics] {10.1002/ctpp.201010017}, \href
  {https://ui.adsabs.harvard.edu/abs/2010CoPP...50...82P} {50, 82}

\bibitem[\protect\citeauthoryear{{Press}, {Teukolsky}, {Vetterling}  \&
  {Flannery}}{{Press} et~al.}{1992}]{Press1992}
{Press} W.~H.,  {Teukolsky} S.~A.,  {Vetterling} W.~T.,   {Flannery} B.~P.,
  1992, {Numerical recipes in C. The art of scientific computing}.
Cambridge Univ. Press

\bibitem[\protect\citeauthoryear{{Rao}, {Meynet}, {Eggenberger},
  {Haemmerl{\'e}}, {Privitera}, {Georgy}, {Ekstr{\"o}m}  \& {Mordasini}}{{Rao}
  et~al.}{2018}]{Rao2018}
{Rao} S.,  {Meynet} G.,  {Eggenberger} P.,  {Haemmerl{\'e}} L.,  {Privitera}
  G.,  {Georgy} C.,  {Ekstr{\"o}m} S.,   {Mordasini} C.,  2018, \mn@doi [\aap]
  {10.1051/0004-6361/201833107}, \href
  {https://ui.adsabs.harvard.edu/abs/2018A&A...618A..18R} {618, A18}

\bibitem[\protect\citeauthoryear{{Rasio}, {Tout}, {Lubow}  \& {Livio}}{{Rasio}
  et~al.}{1996}]{Rasio1996}
{Rasio} F.~A.,  {Tout} C.~A.,  {Lubow} S.~H.,   {Livio} M.,  1996, \mn@doi
  [\apj] {10.1086/177941}, \href
  {https://ui.adsabs.harvard.edu/abs/1996ApJ...470.1187R} {470, 1187}

\bibitem[\protect\citeauthoryear{{Reimers}}{{Reimers}}{1975}]{Reimers1975}
{Reimers} D.,  1975, Memoires of the Societe Royale des Sciences de Liege,
  \href {https://ui.adsabs.harvard.edu/abs/1975MSRSL...8..369R} {8, 369}

\bibitem[\protect\citeauthoryear{{Rein} \& {Liu}}{{Rein} \&
  {Liu}}{2012}]{Rein2012}
{Rein} H.,  {Liu} S.~F.,  2012, \mn@doi [\aap] {10.1051/0004-6361/201118085},
  \href {https://ui.adsabs.harvard.edu/abs/2012A&A...537A.128R} {537, A128}

\bibitem[\protect\citeauthoryear{{Rein} \& {Spiegel}}{{Rein} \&
  {Spiegel}}{2015}]{IAS15}
{Rein} H.,  {Spiegel} D.~S.,  2015, \mn@doi [\mnras] {10.1093/mnras/stu2164},
  \href {https://ui.adsabs.harvard.edu/abs/2015MNRAS.446.1424R} {446, 1424}

\bibitem[\protect\citeauthoryear{{Rein} \& {Tamayo}}{{Rein} \&
  {Tamayo}}{2015}]{WHFAST}
{Rein} H.,  {Tamayo} D.,  2015, \mn@doi [\mnras] {10.1093/mnras/stv1257}, \href
  {https://ui.adsabs.harvard.edu/abs/2015MNRAS.452..376R} {452, 376}

\bibitem[\protect\citeauthoryear{Rein \& Tamayo}{Rein \&
  Tamayo}{2016}]{Rein2016}
Rein H.,  Tamayo D.,  2016, Monthly Notices of the Royal Astronomical Society,
  459, 2275

\bibitem[\protect\citeauthoryear{Rein \& Tamayo}{Rein \& Tamayo}{2017}]{JANUS}
Rein H.,  Tamayo D.,  2017, \mn@doi [MNRAS] {10.1093/mnras/stx2479}, 473, 3351

\bibitem[\protect\citeauthoryear{{Rein} et~al.,}{{Rein}
  et~al.}{2019a}]{Rein2019}
{Rein} H.,  et~al., 2019a, \mn@doi [\mnras] {10.1093/mnras/stz769}, \href
  {https://ui.adsabs.harvard.edu/abs/2019MNRAS.485.5490R} {485, 5490}

\bibitem[\protect\citeauthoryear{Rein, Tamayo  \& Brown}{Rein
  et~al.}{2019b}]{Rein2019b}
Rein H.,  Tamayo D.,   Brown G.,  2019b, Monthly Notices of the Royal
  Astronomical Society, 489, 4632

\bibitem[\protect\citeauthoryear{{Rogers} \& {Nayfonov}}{{Rogers} \&
  {Nayfonov}}{2002}]{Rogers2002}
{Rogers} F.~J.,  {Nayfonov} A.,  2002, \mn@doi [\apj] {10.1086/341894}, \href
  {https://ui.adsabs.harvard.edu/abs/2002ApJ...576.1064R} {576, 1064}

\bibitem[\protect\citeauthoryear{Sackmann, Boothroyd  \& Kraemer}{Sackmann
  et~al.}{1993}]{Sackmann1993}
Sackmann I.-J.,  Boothroyd A.~I.,   Kraemer K.~E.,  1993, ApJ, 418, 457

\bibitem[\protect\citeauthoryear{{Salpeter}}{{Salpeter}}{1954}]{Salpeter1954}
{Salpeter} E.~E.,  1954, \mn@doi [Australian Journal of Physics]
  {10.1071/PH540373}, \href
  {https://ui.adsabs.harvard.edu/\#abs/1954AuJPh...7..373S} {7, 373}

\bibitem[\protect\citeauthoryear{{Saumon}, {Chabrier}  \& {van Horn}}{{Saumon}
  et~al.}{1995}]{Saumon1995}
{Saumon} D.,  {Chabrier} G.,   {van Horn} H.~M.,  1995, \mn@doi [\apjs]
  {10.1086/192204}, \href
  {https://ui.adsabs.harvard.edu/abs/1995ApJS...99..713S} {99, 713}

\bibitem[\protect\citeauthoryear{{Schr{\"o}der} \& {Cuntz}}{{Schr{\"o}der} \&
  {Cuntz}}{2005}]{Schroder2005}
{Schr{\"o}der} K.~P.,  {Cuntz} M.,  2005, \mn@doi [\apjl] {10.1086/491579},
  \href {https://ui.adsabs.harvard.edu/abs/2005ApJ...630L..73S} {630, L73}

\bibitem[\protect\citeauthoryear{{Schr{\"o}der} \& Smith}{{Schr{\"o}der} \&
  Smith}{2008}]{Schroder2008}
{Schr{\"o}der} K.-P.,  Smith R.~C.,  2008, \mn@doi [\mnras]
  {10.1111/j.1365-2966.2008.13022.x}, \href
  {https://ui.adsabs.harvard.edu/abs/2008MNRAS.386..155S} {386, 155}

\bibitem[\protect\citeauthoryear{Strang}{Strang}{1968}]{Strang1968}
Strang G.,  1968, SIAM Journal on Numerical Analysis, 5, 506

\bibitem[\protect\citeauthoryear{{Tamayo}, {Rein}, {Shi}  \&
  {Hernandez}}{{Tamayo} et~al.}{2020}]{Tamayo2020}
{Tamayo} D.,  {Rein} H.,  {Shi} P.,   {Hernandez} D.~M.,  2020, \mn@doi
  [\mnras] {10.1093/mnras/stz2870}, \href
  {https://ui.adsabs.harvard.edu/abs/2020MNRAS.491.2885T} {491, 2885}

\bibitem[\protect\citeauthoryear{{Timmes} \& {Swesty}}{{Timmes} \&
  {Swesty}}{2000}]{Timmes2000}
{Timmes} F.~X.,  {Swesty} F.~D.,  2000, \mn@doi [\apjs] {10.1086/313304}, \href
  {https://ui.adsabs.harvard.edu/abs/2000ApJS..126..501T} {126, 501}

\bibitem[\protect\citeauthoryear{{Vassiliadis} \& {Wood}}{{Vassiliadis} \&
  {Wood}}{1993}]{Vassiliadis1993}
{Vassiliadis} E.,  {Wood} P.~R.,  1993, \mn@doi [\apj] {10.1086/173033}, \href
  {https://ui.adsabs.harvard.edu/abs/1993ApJ...413..641V} {413, 641}

\bibitem[\protect\citeauthoryear{Veras}{Veras}{2016a}]{Veras2016a}
Veras D.,  2016a, \mn@doi [Royal Society Open Science] {10.1098/rsos.150571},
  3, 150571

\bibitem[\protect\citeauthoryear{Veras}{Veras}{2016b}]{Veras2016b}
Veras D.,  2016b, \mn@doi [Monthly Notices of the Royal Astronomical Society]
  {10.1093/mnras/stw2170}, 463, 2958

\bibitem[\protect\citeauthoryear{Veras \& Wyatt}{Veras \&
  Wyatt}{2012}]{Veras2012}
Veras D.,  Wyatt M.~C.,  2012, \mn@doi [MNRAS]
  {10.1111/j.1365-2966.2012.20522.x}, 421, 2969

\bibitem[\protect\citeauthoryear{Veras, Mustill, Bonsor  \& Wyatt}{Veras
  et~al.}{2013}]{Veras2013}
Veras D.,  Mustill A.~J.,  Bonsor A.,   Wyatt M.~C.,  2013, \mn@doi [Monthly
  Notices of the Royal Astronomical Society] {10.1093/mnras/stt289}, 431, 1686

\bibitem[\protect\citeauthoryear{Villaver, Livio, Mustill  \& Siess}{Villaver
  et~al.}{2014}]{Villaver}
Villaver E.,  Livio M.,  Mustill A.~J.,   Siess L.,  2014, \mn@doi [The
  Astrophysical Journal] {10.1088/0004-637x/794/1/3}, 794, 3

\bibitem[\protect\citeauthoryear{{Zahn}}{{Zahn}}{1977}]{Zahn1977}
{Zahn} J.~P.,  1977, \aap, \href
  {https://ui.adsabs.harvard.edu/abs/1977A&A....57..383Z} {500, 121}

\bibitem[\protect\citeauthoryear{{Zahn}}{{Zahn}}{1989}]{Zahn1989}
{Zahn} J.~P.,  1989, \aap, \href
  {https://ui.adsabs.harvard.edu/abs/1989A&A...220..112Z} {220, 112}

\makeatother
\end{thebibliography}

%\begin{thebibliography}{99}
%\end{thebibliography}

%%%%%%%%%%%%%%%%%%%%%%%%%%%%%%%%%%%%%%%%%%%%%%%%%%

%%%%%%%%%%%%%%%%% APPENDICES %%%%%%%%%%%%%%%%%%%%%

\appendix

%%%%%%%%%%%%%%%%%%%%%%%%%%%%%%%%%%%%%%%%%%%%%%%%%%

% Don't change these lines
\bsp	% typesetting comment
\label{lastpage}
\end{document}